\documentclass[aps,pra,twocolumn,superscriptaddress,nofootinbib,10pt]{revtex4-2}
\usepackage{graphicx}
\usepackage{amsmath,amssymb}
\usepackage{bm}
\usepackage{color}

\begin{document}

\title{Quantum and classical entropic complexity of the thermal state: coherence, decoherence,
and the ergodic-to-localized crossover in random-matrix and many-body models}

\author{Imre Varga}
\affiliation{Department of Theoretical Physics, Institute of Physics, Budapest University of
Technology and Economics, M\H{u}egyetem rkp.\ 3., H-1111 Budapest, Hungary}

\author{Onur Aktan}
\affiliation{Institute of Mathematics, Budapest University of Technology and Economics,
M\H{u}egyetem rkp.\ 3., H-1111 Budapest, Hungary}

\date{\today}

\begin{abstract}
Does thermal averaging preserve signatures of eigenstate complexity? We study the entropic complexity
$C=S_1-S_2$ (Shannon minus second-order R\'enyi entropy) across the ergodic-to-localized crossover of
three disordered models -- the Rosenzweig--Porter (RP) ensemble, the power-law banded random matrix
(PLBRM) ensemble, and the random-field Heisenberg chain -- comparing a purely wavefunction-level
quantity, $C_{\rm eig}$, to the complexity of the thermal (Gibbs) state itself, evaluated before and
after pointer-basis dephasing via the {\it trace} ($C_{\rm tr}$) and {\it diagonal} ($C_{\rm diag}$)
complexities. The answer is mostly no: thermal averaging strongly suppresses, but does not eliminate,
eigenstate-complexity signatures. $C_{\rm eig}$ develops a pronounced mid-phase maximum in RP
($\gamma^\ast\approx1.5$--$1.6$) and, at a matching fractal dimension, in the structurally independent
PLBRM model ($b^\ast=0.31\pm0.01$, independent of $N=200$--$1600$); a high-statistics PLBRM scan resolves
a weak ($\approx10\%$) but reproducible thermal shadow of this peak in $\max_T C_{\rm diag}(b)$, showing
the suppression is strong but not absolute. This mid-phase eigenstate feature is confined to the two
random-matrix models, which possess an extended intermediate (multifractal) phase; in the interacting
random-field Heisenberg chain, whose non-ergodic extended states are instead confined to the critical
region of a single ergodic-to-localized crossover, the eigenstate complexity is known to peak {\it at}
that many-body localization transition ($W_c/J\approx3.5$--$4$) rather than mid-phase, so we use that
model to probe the thermal-state side of the comparison. A second, genuinely thermal feature -- an {\it edge} just
inside the ergodic phase ($\gamma\approx0.8$--$0.9$ in RP, $W^\ast/J\approx0.5$--$0.7$ in the Heisenberg
chain) -- has no eigenstate-level counterpart and, in PLBRM, recedes with $N$ (approximately
$b_{\rm edge}^\ast\propto N^{0.60\pm0.07}$) even though the mid-phase location does not. Both features are
tracked by simple, scale-invariant crossover indicators: the log-ratio
$\mathcal L=\ln(T^\ast_{\rm tr}/T^\ast_{\rm diag})$, which vanishes upon localization in all three
models, and the relative entropy of coherence $C_{\rm rel}$, a genuine coherence monotone whose
normalized $T/W_{\rm eff}$ profile orders by density-of-states class but does not collapse to a universal
curve. Taken together, entropic complexity
cleanly separates thermal-state and eigenstate physics: a sharp wavefunction-level feature, reproducible
across unrelated random-matrix constructions, leaves only a faint and structurally distinct imprint on
thermal observables.
\end{abstract}

\maketitle

\section{Introduction}

Disordered quantum systems interpolate between chaotic, ergodic behavior and integrable or localized
behavior as a control parameter -- disorder strength, coupling, or a tuning exponent -- is varied
\cite{Vojta2019,GornyiMirlin2002,AbaninColloquium2019}. Random matrix theory has long provided the
minimal, analytically tractable setting in which to organize this phenomenology
\cite{BohigasGiannoniSchmit1984,Haake1991}, and the Rosenzweig--Porter (RP) ensemble
\cite{RosenzweigPorter1960,KravtsovKhaymovichCuevasAmini2015} has become the canonical toy model for
a genuine intermediate phase: a regime of non-ergodic but extended (multifractal) eigenstates,
sandwiched between an ergodic Wigner--Dyson phase and an Anderson-localized, Poissonian one. This
same phenomenology -- ergodic, intermediate, localized -- is believed to be relevant, in some form,
to the many-body-localization (MBL) transition of interacting disordered spin chains
\cite{BaskoAleinerAltshuler2006,AbaninColloquium2019}, although the microscopic mechanism and the
degree to which random-matrix universality survives interactions remain active questions
\cite{SuntajsBoncaProsenVidmar2019}.

A recurring difficulty in this program is that different diagnostics of ``complexity'' or
``ergodicity breaking'' do not agree on where, or how sharply, a system crosses over between regimes.
The fractal dimension $D_2$, extracted from the inverse participation ratio, is the traditional
observable, but recent work has shown that other markers -- entanglement entropy, non-stabilizerness
(``magic''), and coherence-based measures -- can detect transitions at parametrically different
points once a genuine intermediate phase is present, while collapsing onto a single transition point
when it is absent \cite{SantraWindeyBandyopadhyayLegramandiHauke2025}. A closely related entropic
complexity measure, $S_C=S-R_2$, built from exactly the Shannon and second-order R\'enyi entropies
used here, has previously been used to locate the crossover from integrable/localized to
quantum-chaotic/ergodic behavior in deformed random-matrix ensembles and the disordered Heisenberg
chain \cite{Varga2026}, with the maximum complexity found at the boundary between order and chaos.
This raises the question this paper is built around: {\it does thermal averaging preserve signatures of
eigenstate complexity}, or does coupling a complexity measure to a thermal environment wash them out?

This paper is organized around that same functional, the entropic complexity $C=S_1-S_2$, the
difference between the Shannon entropy and the second-order R\'enyi entropy of a probability
distribution over a preferred (site, or ``pointer'') basis. It vanishes for both extreme uniform
(fully delocalized) and extreme concentrated (fully localized) distributions, and is maximal for
distributions of intermediate structure, making it a natural, LMC-style measure of ``interesting''
intermediate statistics \cite{LopezRuizManciniCalbet1995,CatalanGarayAngulo2002}. We use it in two
distinct ways: as a property of individual, pure energy eigenstates, $C_{\rm eig}=\langle C\rangle$
averaged over a bulk window of the spectrum, and as a property of the thermal (Gibbs) state
$\rho_T=e^{-H/T}/Z$, evaluated either from the eigenvalue populations alone (the trace complexity
$C_{\rm tr}$, manifestly basis-independent and unitarily invariant, built entirely from $\rho_T$'s
eigenvalues and hence free of any pointer-basis choice) or from the diagonal populations in the site
basis after the state has been completely dephased (the diagonal complexity $C_{\rm diag}$, built
from the classical probability distribution that survives complete dephasing in the pointer basis).
Raising $T$ from zero changes the Boltzmann occupation of the energy eigenstates; it does not, by
itself, destroy any coherence. What isolates the coherence is instead the comparison of $\rho_T$ to
its pointer-basis-dephased counterpart at fixed $T$: the gap $C_{\rm gap}=C_{\rm diag}-C_{\rm tr}$
quantifies, at each disorder strength and temperature, how much of the thermal-state complexity is
carried by pointer-basis coherence rather than by the eigenvalue populations alone. The whole study is
organized by the ergodic-to-localized crossover of these disordered models, and by how $C_{\rm tr}$ and
$C_{\rm diag}$ -- and their characteristic temperatures -- track it. A closely related,
rigorously monotone coherence measure, $C_{\rm rel}=S(\rho_{\rm diag})-S(\rho)$, the relative entropy
of coherence \cite{BaumgratzCramerPlenio2014}, is obtained by keeping only the Shannon ($S_1$) parts
of this construction.

We apply this framework to three models: the generalized RP ensemble, the power-law banded random
matrix (PLBRM) ensemble \cite{MirlinFyodorovDittesQuezadaSeligman1996}, and the random-field Heisenberg
spin-$\tfrac12$ chain, the standard microscopic model for MBL
\cite{PalHuse2010,LuitzLaflorencieAlet2015}. Our central empirical finding, stated plainly, is this:
{\it thermal averaging suppresses, but does not eliminate, eigenstate-complexity signatures}. Concretely,
these models support (at least) two structurally distinct complexity features that must not be
conflated: an {\it edge} feature, located just inside the ergodic/thermal phase, visible in genuinely
thermal (mixed-state) diagnostics and connected to ground-state/low-temperature physics; and a
{\it mid-phase} feature, located deep inside the fractal/intermediate regime, which is predominantly a
single-eigenstate wavefunction statistic. The mid-phase feature is largely a property of $C_{\rm eig}$
alone, strongly suppressed by thermal or microcanonical mixing -- though, in PLBRM's higher-statistics
data, not entirely erased -- while the edge feature is a property of the Gibbs state $\rho_T$ (via
$C_{\rm tr}$, $C_{\rm diag}$, and their difference) and has no counterpart at the single-eigenstate
level. In both RP and PLBRM the mid-phase feature sits at a fixed point in the tuning parameter
($\gamma^\ast$ or $b^\ast$), independent of $N$; the edge feature, by contrast, recedes with $N$ in
PLBRM even though the mid-phase feature in the same model does not, showing that the two features are
governed by distinct, only weakly coupled mechanisms even within a single model. Establishing this
distinction, quantifying the thermal edge feature across all three models and the mid-phase eigenstate
feature in the two random-matrix models that possess an extended fractal phase, and identifying what is still missing
to complete the comparison, is the content of this paper.

\section{Models and complexity measures}

\subsection{Entropic complexity of a distribution}
For a normalized probability distribution $\{p_i\}_{i=1}^{L}$ over $L$ basis states we define the
Shannon entropy $S_1=-\sum_i p_i\ln p_i$ and the second-order R\'enyi entropy
$S_2=-\ln\sum_i p_i^2=-\ln(\mathrm{IPR}_2)$, and the entropic complexity
\begin{equation}
C \;=\; S_1-S_2 \;\geq 0,
\end{equation}
with equality only for exactly flat or exactly one-hot distributions. For a pure state
$|\psi\rangle=\sum_i\psi_i|i\rangle$ we take $p_i=|\psi_i|^2$ and write $C_{\rm eig}$ for the
disorder- and spectrum-window average of $C[\psi]$ over energy eigenstates.

\subsection{Thermal state, trace/diagonal complexity, coherence gap}
\begin{figure*}[t]
\includegraphics[width=\textwidth]{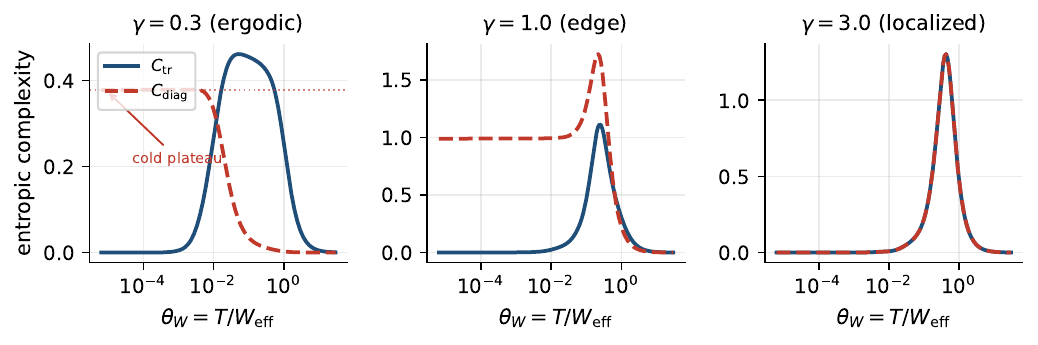}
\caption{Temperature dependence of the trace ($C_{\rm tr}$, solid) and diagonal ($C_{\rm diag}$,
dashed) complexities, illustrated for the RP model ($N=1600$) at three values of $\gamma$. At
$\gamma=3$ (localized) both curves are single, coincident, broadened peaks. At $\gamma=1$ (edge)
$C_{\rm diag}$ develops a peak above a residual low-$T$ shoulder. At $\gamma=0.3$ (ergodic)
$C_{\rm diag}$ has {\it no} interior maximum: it rises to a cold plateau as $T\to0$ (the ground-state
site complexity) while $C_{\rm tr}$ remains a clean interior peak. Temperature is in bandwidth units
$\theta_W=T/W_{\rm eff}$.}
\label{fig:ecshapes}
\end{figure*}

For the Gibbs state $\rho_T=e^{-H/T}/Z=\sum_n p_n(T)|n\rangle\langle n|$ ($|n\rangle$ energy
eigenstates, $p_n=e^{-E_n/T}/Z$), the {\it trace} complexity is computed from the eigenvalues alone,
\begin{equation}
C_{\rm tr}(T) = C[\{p_n(T)\}],
\end{equation}
a purely thermodynamic, basis-independent quantity that carries no eigenvector information. The
{\it diagonal} complexity is computed from the site-basis populations
$\rho_{ii}(T)=\sum_n p_n(T)|\psi_n(i)|^2$,
\begin{equation}
C_{\rm diag}(T) = C[\{\rho_{ii}(T)\}],
\end{equation}
which does depend on the eigenvectors. Both interpolate between the same two limits,
$C_{\rm tr,diag}(T\!\to\!\infty)=0$ ($\rho\to \mathbb{1}/L$, structureless in every basis) and a
finite $T\!\to\!0$ value set by the ground state, and are typically non-monotonic in $T$,
defining characteristic temperatures $T^\ast_{\rm tr}$, $T^\ast_{\rm diag}$ at their maxima
(Fig.~\ref{fig:ecshapes}). An important exception, central below, is that $C_{\rm diag}(T)$ can
{\it lose} its interior maximum and instead saturate to a {\it cold plateau} as $T\to0$: when the
eigenvectors are strongly delocalized in the site basis, the ground-state site distribution is
already highly structured, so $C_{\rm diag}$ is largest at $T\to0$ and simply decreases with
temperature -- there is then no well-defined $T^\ast_{\rm diag}$ (Fig.~\ref{fig:ecshapes}, left). The
trace complexity, by contrast, always vanishes at $T\to0$ (a single dominant eigenvalue) and thus
retains a genuine interior peak.

The coherence gap
\begin{equation}
C_{\rm gap}(T) = C_{\rm diag}(T) - C_{\rm tr}(T)
\end{equation}
measures the extra complexity that appears in the pointer basis relative to the intrinsic spectral
one. It vanishes at $T\to\infty$ (full decoherence) and equals the (nonnegative) ground-state
site-basis complexity at $T\to0$; at intermediate $T$ it is {\it not} sign-definite -- it can become
negative when strongly delocalized eigenvectors drive $\rho_{ii}$ toward uniform, suppressing
$C_{\rm diag}$ below $C_{\rm tr}$ (Sec.~\ref{sec:resource}). Its temperature maximum, $\max_T
C_{\rm gap}\geq0$, is the quantity we use as a transition marker. We also use the rigorous
coherence monotone obtained from the Shannon parts alone,
$C_{\rm rel}(\rho)=S_1[\{\rho_{ii}\}]-S_1[\{p_n\}]=S(\rho_{\rm diag}\|\rho)$, the relative entropy of
coherence \cite{BaumgratzCramerPlenio2014}, which is non-increasing under every incoherent operation
and, unlike $C_{\rm gap}$, provably cannot be enhanced by pure dephasing.

\subsection{Three coherence measures and the $\Delta S_2=D_2\ln N$ identity}
\label{sec:three_coherence}
It is useful to separate the Shannon and R\'enyi-2 content of the coherence gap explicitly. Writing
$S_1^{\rm diag}=S_1[\{\rho_{ii}\}]$, $S_1^{\rm full}=S_1[\{p_n\}]=S(\rho)$ (the von Neumann entropy),
$S_2^{\rm diag}=-\ln\sum_i\rho_{ii}^2$ and $S_2^{\rm full}=-\ln\mathrm{Tr}\,\rho^2=-\ln\sum_n p_n^2$,
the gap decomposes as
\begin{equation}
C_{\rm gap} = C_{\rm diag}-C_{\rm tr}
= \underbrace{(S_1^{\rm diag}-S_1^{\rm full})}_{C_{\rm rel}}
-\underbrace{(S_2^{\rm diag}-S_2^{\rm full})}_{\Delta S_2},
\label{eq:decomp}
\end{equation}
i.e. $C_{\rm gap}=C_{\rm rel}-\Delta S_2$. The Shannon part $C_{\rm rel}=S(\rho_{\rm diag})-S(\rho)$ is
the relative entropy of coherence \cite{BaumgratzCramerPlenio2014}, a genuine coherence monotone. The
R\'enyi-2 part $\Delta S_2=S_2^{\rm diag}-S_2^{\rm full}\geq0$ (nonnegative because dephasing cannot
increase purity, $\mathrm{Tr}\,\rho^2\geq\sum_i\rho_{ii}^2$) is the historically first decoherence
diagnostic used for these models: it vanishes for a fully dephased (localized) state and is maximal
in the ergodic phase. Its zero-temperature limit is fixed by the ground-state wavefunction: since
$S_2^{\rm full}(T\!\to\!0)\to0$ (pure state) while $S_2^{\rm diag}(T\!\to\!0)\to-\ln\mathrm{IPR}_2[\psi_0]$,
\begin{equation}
\frac{\Delta S_2(T\!\to\!0)}{\ln N}\;\longrightarrow\;D_2,
\label{eq:coldD2}
\end{equation}
the fractal dimension of the ground state. Thus $\Delta S_2$ is a mixed-state, thermally-defined probe
that reduces to the eigenstate fractal dimension in the cold limit. The three measures play
complementary roles, made explicit in Sec.~\ref{sec:resource}: $C_{\rm rel}$ is the rigorous resource
monotone (monotone in the control parameter, monotone in $T$, non-increasing under dephasing);
$\Delta S_2$ carries the $D_2$ identity (\ref{eq:coldD2}); and $C_{\rm gap}=C_{\rm rel}-\Delta S_2$ is
neither sign-definite nor monotone but, maximized over $T$, is a sharp transition marker.

\subsection{Rosenzweig--Porter model}
The generalized RP Hamiltonian is
\begin{equation}
H_{ij} = h_i\,\delta_{ij} + N^{-\gamma/2} V_{ij}(1-\delta_{ij}),
\label{eq:rp}
\end{equation}
with $h_i$ i.i.d.\ Gaussian on-site disorder and $V$ a GOE matrix with unit off-diagonal variance.
This model has two exact transitions in the thermodynamic limit: an ergodic-to-fractal transition at
$\gamma=1$ and a fractal-to-localized (Anderson) transition at $\gamma=2$, with a mono-fractal
dimension $D_q=2-\gamma$ for $1\le\gamma\le2$, independent of $q$
\cite{KravtsovKhaymovichCuevasAmini2015}. Level statistics (mean gap ratio $\langle r\rangle$) track
only the second transition, remaining at the GOE value throughout the fractal phase
\cite{Facoetti2016,BogomolnySieber2018}, the well-known decoupling of spectral and eigenvector
transitions in this model.

\subsection{Power-law banded random matrix model}
\label{sec:plbrm_model}
We separately consider the power-law banded random matrix (PLBRM) ensemble
\cite{MirlinFyodorovDittesQuezadaSeligman1996}, an $N\times N$ real symmetric matrix with independent
Gaussian entries of variance
\begin{equation}
\langle H_{ij}^2\rangle = \big[1+(d_{ij}/b)^{2\mu}\big]^{-1},\qquad i\neq j,
\label{eq:plbrm}
\end{equation}
where $d_{ij}=\min(|i-j|,N-|i-j|)$ is the distance between sites $i,j$ arranged on a ring, $b$ a
bandwidth parameter, and $H_{ii}$ i.i.d.\ $\mathcal N(0,1)$ on-site disorder. At the critical exponent
$\mu=1$ the model is multifractal for {\it every} value of $b$: a nonlinear-$\sigma$-model mapping
finds a continuous family of critical theories parametrized by $b$
\cite{MirlinFyodorovDittesQuezadaSeligman1996}, interpolating between an Anderson-localized limit
($b\to0$) and a Gaussian-orthogonal-ensemble limit ($b\to\infty$), with both the fractal dimension and
the level statistics varying continuously with $b$. We fix $\mu=1$ and use $b$ as the tuning
parameter, in direct analogy to $\gamma$ in RP. Unlike RP, PLBRM has no two-transition structure in its
tuning parameter -- it is critical along the entire line -- so any peak found in $C_{\rm eig}(b)$
cannot be attributed to proximity to a phase boundary in the RP sense, only to the multifractal
wavefunction statistics themselves. PLBRM has recently been proposed, independently of the present
work, as an effective single-particle proxy for ergodicity breaking in interacting many-body
Hamiltonians \cite{BuijsmanHaqueKhaymovich2026,CohenOzZhong2024}; we use it here as a second,
structurally independent multifractal random-matrix model -- translationally invariant and
power-law-banded rather than RP's all-to-all coupling -- to test whether the mid-phase entropic-
complexity feature identified in RP (Sec.~\ref{sec:rp_eig}) is a peculiarity of the RP construction or
a more general property of multifractal wavefunction statistics.

\subsection{Random-field Heisenberg chain}
The third model is the standard MBL testbed,
\begin{equation}
H = J\sum_{i=1}^{L-1}\bm S_i\cdot\bm S_{i+1} + \sum_{i=1}^{L} h_i S_i^z,\qquad h_i\in[-W,W],
\label{eq:heis}
\end{equation}
open boundary conditions, exact-diagonalized for $L=10,12,14,16$. This model has a numerically
well-established ergodic-to-MBL crossover at $W_c/J\approx3.5$--$4$ in this size range
\cite{LuitzLaflorencieAlet2015}, with the caveat that the precise location, and even the existence of
a sharp transition in the thermodynamic limit, remains debated
\cite{SuntajsBoncaProsenVidmar2019,AbaninColloquium2019}. The site (computational, $S^z$-product)
basis plays the role of the pointer basis for $C_{\rm diag}$ and $C_{\rm gap}$.

\section{Results}

\begin{figure*}[t]
\includegraphics[width=\textwidth]{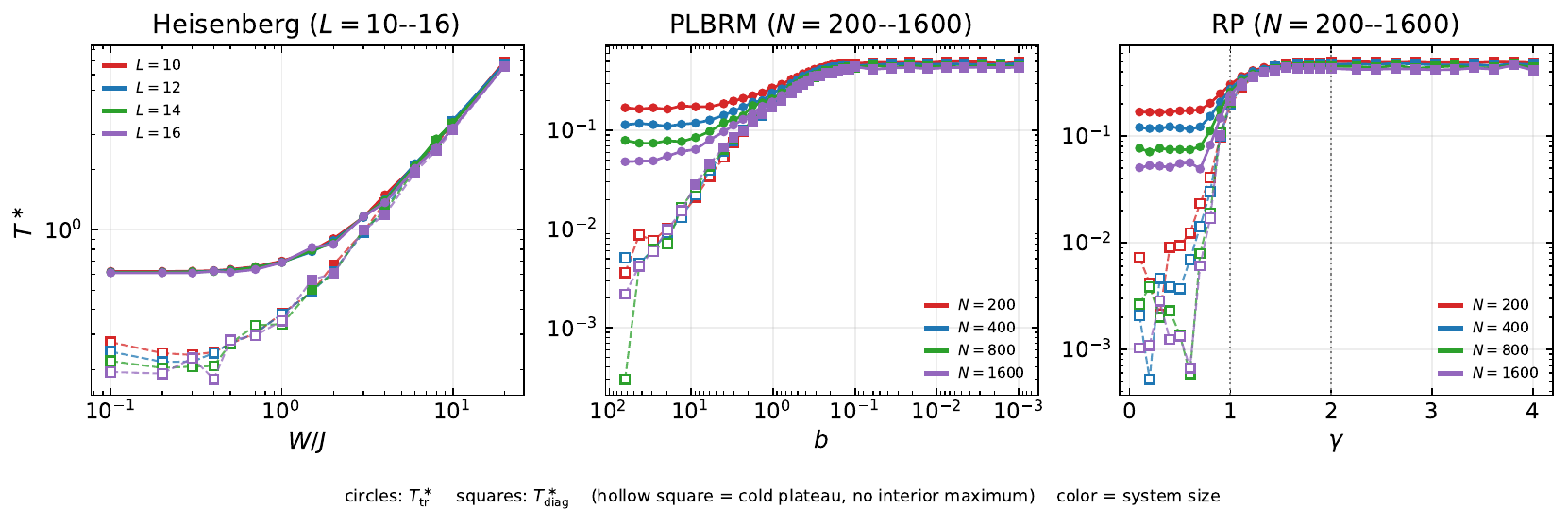}
\caption{Trace versus diagonal characteristic temperatures across the three models and their full
range of system sizes: Heisenberg ($L=10$--$16$) vs disorder $W$, PLBRM ($N=200$--$1600$) vs bandwidth
$b$, and RP ($N=200$--$1600$) vs $\gamma$, with color marking size (red, blue, green, violet for the
smallest to largest size in each model, as in Figs.~\ref{fig:plbrm} and \ref{fig:plbrm_thermal}) and
marker shape marking the two temperatures ($T^\ast_{\rm tr}$: circles, $T^\ast_{\rm diag}$: squares).
The $b$ axis in the center panel is reversed (increasing rightward, toward the PLBRM localized limit)
so that all three panels share the same left-to-right ergodic-to-localized direction. In each case
$T^\ast_{\rm tr}$ and $T^\ast_{\rm diag}$ {\it coincide on the localized side} (Heisenberg large $W$,
PLBRM small $b$, RP large $\gamma$), where the pointer basis is the eigenbasis and all sizes collapse
onto a single curve, and {\it separate on the ergodic side}, where $T^\ast_{\rm diag}<T^\ast_{\rm tr}$
or -- deeper into the ergodic regime (hollow squares) -- $C_{\rm diag}$ becomes a cold plateau with no
interior maximum at all, so that $T^\ast_{\rm diag}$ is undefined. The size dependence itself is
visibly asymmetric: $T^\ast_{\rm tr}$ is essentially size-invariant for the Heisenberg chain at every
$W$, while in both random-matrix models it decreases systematically with $N$ on the ergodic side before
collapsing with the other sizes on the localized side -- consistent with $T^\ast_{\rm tr}$ tracking an
intensive energy scale in the many-body model but an $N$-dependent bandwidth-normalized scale in RP and
PLBRM. The vertical dotted lines in the RP panel mark $\gamma=1,2$.}
\label{fig:tstar_compare}
\end{figure*}

The results below establish one claim, in stages: entropic complexity cleanly separates
eigenstate-level from thermal-state-level physics, and thermal averaging suppresses -- strongly, but
not completely -- the eigenstate-level signature. Figure~\ref{fig:tstar_compare} previews the pattern
that recurs in every model below: trace and diagonal characteristic temperatures coincide on the
localized side and separate on the ergodic side, in RP, PLBRM, and the Heisenberg chain alike.
Sections~\ref{sec:rp_eig}--\ref{sec:rp_coherence} establish the distinction in the RP model, where it
is cleanest to isolate; Sec.~\ref{sec:plbrm} is the paper's central cross-check, reproducing both the
mid-phase eigenstate feature and, at an order of magnitude higher statistics, its weak thermal shadow,
in a second, structurally independent random-matrix model; Sec.~\ref{sec:heis} extends the comparison
to the genuinely many-body, interacting setting; Sec.~\ref{sec:resource} isolates the rigorous
coherence-resource content of the thermal-state side; and Sec.~\ref{sec:synthesis} draws the three
models together.

\subsection{RP: the eigenstate complexity peaks inside the fractal phase}
\label{sec:rp_eig}

\begin{figure*}[t]
\includegraphics[width=\textwidth]{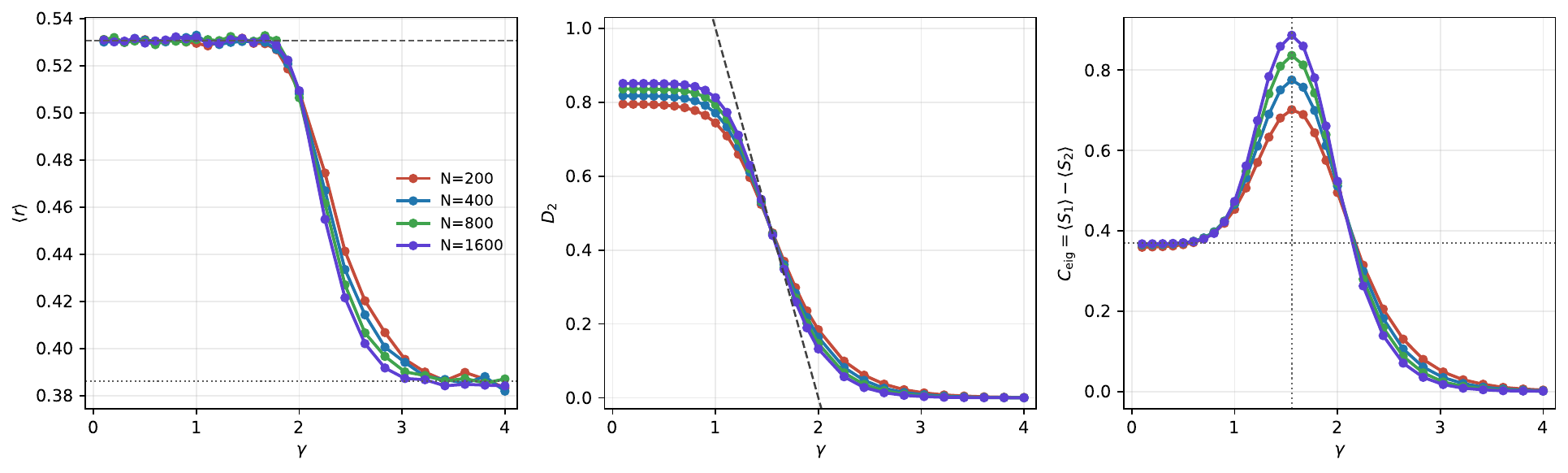}
\caption{Rosenzweig--Porter model, $N=200$--$1600$. Left: mean gap ratio $\langle r\rangle(\gamma)$,
pinned to the GOE value (dashed) throughout the ergodic and fractal phases and falling toward Poisson
(dotted) only past $\gamma=2$. Center: fractal dimension $D_2(\gamma)$, tracking $2-\gamma$ (dashed)
with finite-size rounding at the boundaries. Right: eigenstate entropic complexity $C_{\rm
eig}(\gamma)$, peaking at $\gamma^\ast\approx1.5$--$1.6$ (vertical dotted line at $\gamma^\ast\approx1.56$);
the horizontal dotted line marks the analytic GOE plateau
$C_{\rm GOE}=\ln\frac32-\psi(\frac32)\approx0.369$.}
\label{fig:rp_markers}
\end{figure*}

For a single eigenstate in the Breit--Wigner (single-resonance) approximation, the local intensity
profile is a Lorentzian of width $\Gamma\sim N^{1-\gamma}$ centered at the eigenvalue, and the
R\'enyi moments can be computed in closed form,
$\mathrm{IPR}_q \simeq \mathcal C(q)\,(N\Gamma)^{1-q}$ with
$\mathcal C(q)=2^{q-1}\pi^{1/2-q}\Gamma(q-\tfrac12)/\Gamma(q)$ (not to be confused with the
complexity $C$), giving $S_q = D\ln N + f(q)$ with $f(q)=-\ln\mathcal C(q)/(q-1)$ and $D=2-\gamma$.
The leading, extensive $D\ln N$ piece cancels identically in $C_{\rm eig}=S_1-S_2$, leaving a
mean-field prediction that is {\it flat} across the entire fractal phase,
\begin{equation}
C_{\rm eig}^{\rm MF} = f(1)-f(2) = \ln(2\pi)-\ln\pi = \ln2 \approx 0.693,
\end{equation}
independent of $\gamma$ and $N$. For comparison, the exact ergodic (GOE) value, obtained from the
Dirichlet($\tfrac12,\ldots,\tfrac12$) statistics of Haar-random eigenvector intensities, is
\begin{equation}
C_{\rm GOE} = \ln\tfrac32 - \psi(\tfrac32) \approx 0.369.
\end{equation}
Neither prediction anticipates a $\gamma$-dependent maximum. Numerically, however, $C_{\rm
eig}(\gamma)$ is not flat: it rises well above $\ln2$ and develops a pronounced maximum inside the
fractal phase (Fig.~\ref{fig:rp_markers}, right), at $\gamma^\ast\approx1.5$--$1.6$ essentially
independent of $N$ over the range studied ($N=200$--$1600$), with only the peak {\it height} growing
slowly with $N$ (from $C_{\rm eig}^\ast\approx0.70$ at $N=200$ to $\approx0.89$ at $N=1600$). A dedicated
high-resolution scan at $N=1600$ ($\Delta\gamma\approx0.03$ over $\gamma\in[1.30,1.70]$, $60$
realizations per point), with the peak location estimated the same way as $b^\ast$ in PLBRM (parabolic
interpolation in $\ln\gamma$, $4000$-draw bootstrap over disorder realizations), gives
$\gamma^\ast=1.561\pm0.008$ at this size -- consistent with, and considerably sharper than, the
$\gamma^\ast\approx1.5$--$1.6$ range quoted above from the coarser production grid. Both the
fractal dimension $D_2(\gamma)$ and the mean gap ratio $\langle r\rangle(\gamma)$
(Fig.~\ref{fig:rp_markers}, center and left) are featureless at $\gamma^\ast$: $D_2$ decreases
smoothly through $2-\gamma$ with no inflection, and $\langle r\rangle$ remains pinned to the GOE value
until $\gamma\approx1.9$--$2$. The peak in $C_{\rm eig}$ is therefore invisible to both the standard
multifractal and the standard spectral diagnostics -- it is a genuine, independent complexity
feature.

A natural hypothesis is that $\gamma^\ast$ is pinned to the point where the mean fractal dimension
crosses one half, $D_2=1/2\Leftrightarrow\gamma=1.5$ exactly, since this is the only special point
the analytic $D_2=2-\gamma$ law singles out. A refined scan with $\Delta\gamma=0.02$ around the peak
does {\it not} support this: at every $N=200$--$1600$ studied, the parabolic
peak of $C_{\rm eig}(\gamma)$ sits $0.06$--$0.10$ above the $\gamma$ at which the {\it measured} $D_2$
actually equals $1/2$; $D_2$ at the $C_{\rm eig}$ maximum is consistently $\approx0.43$--$0.45$, not
$0.5$. Extrapolating both quantities separately in $1/\ln N$ gives suggestively close asymptotic
values ($\gamma^\ast_{\rm peak}(\infty)\approx1.48$, $D_2\!=\!1/2$ crossing $\to\approx1.48$ already at
$N=1600$), but this rests on only four system sizes and a peak-fit that is sensitive to noise on the
rather flat maximum, so we regard the question of whether the two quantities coincide only
asymptotically, or remain offset, as open (Sec.~\ref{sec:voids}). What the refined grid does rule out
is the simplest, size-independent explanation: $D_2=1/2$ is not the mechanism at accessible sizes.

A qualitative account of why the peak nonetheless anchors near $D_2\approx0.44$ follows from reading
$C=S_1-S_2$ as an LMC-type balance between disorder and structure. Two effects compete as $\gamma$
increases across the fractal phase: the relative variance of the site intensities grows -- the
multifractal fluctuations responsible for $C_{\rm eig}$ exceeding its mean-field value $\ln2$
(Appendix~\ref{app:selfavg}) sharpen as the states concentrate -- while the overall spread of the
occupied region shrinks toward the localized, one-hot limit where $C\to0$. The maximum sits where these
balance, and that it falls slightly on the more-localized side of the $D_2=1/2$ point
($\gamma^\ast\approx1.56$, $D_2\approx0.44$) suggests the fluctuation growth still dominates just past
$\gamma=1.5$ before the support shrinkage takes over. We offer this as a qualitative conjecture, to be
made precise by the fluctuation-resolved R\'enyi-moment calculation left open as item (i) of
Sec.~\ref{sec:voids}.

Independent support for a genuine, non-$D_2$ feature near $\gamma\approx1.5$--$1.6$ comes from a
recent study of complexity markers in the RP model \cite{SantraWindeyBandyopadhyayLegramandiHauke2025},
which finds, via finite-size scaling of derivative-crossing points for bulk eigenstates, transition
locations $\gamma_1^\infty[D_2]=1.03\pm0.03$ (matching the analytic value) but
$\gamma_1^\infty[S_{vN}]=1.51\pm0.22$ for the maximal half-chain entanglement entropy and
$\gamma_1^\infty[\mathcal M_2]=1.60\pm0.06$ for the stabilizer R\'enyi entropy (``magic''). These two
markers depart from their ergodic plateau distinctly later than $D_2$, bracketing our $\gamma^\ast$
almost exactly. While $S_{vN}$, $\mathcal M_2$, and $C_{\rm eig}$ are conceptually different
quantities, this is strong circumstantial evidence that $\gamma\approx1.5$--$1.6$ marks a real,
reproducible ``deep-fractal-phase'' crossover that the fractal dimension alone cannot see.

Because $C_{\rm eig}$ is evaluated in the site basis throughout, a natural concern is whether the
$\gamma^\ast$ peak is a property of the disordered Hamiltonian at all, or merely an artifact of
measuring intensity in {\it some particular} basis. We checked this directly: for $N=800$ and the same
bulk eigenstates used above, we recomputed $C_{\rm eig}(\gamma)$ after rotating every eigenvector by an
independent Haar-random orthogonal matrix (a different, freshly drawn rotation for each disorder
realization), so that the ``pointer basis'' used to define site populations is generic rather than the
physical, disorder-diagonal one. The result is unambiguous: the random-basis complexity is flat within
error across the entire range $\gamma=0.3$--$3.0$, $C_{\rm eig}^{\rm rand}=0.366\pm0.001$, statistically
indistinguishable from the analytic GOE value $C_{\rm GOE}\approx0.369$ at every single $\gamma$ studied
-- including deep in the localized phase ($\gamma=3$), where the site-basis value collapses to
$C_{\rm eig}^{\rm site}=0.027\pm0.001$, and at the mid-phase maximum ($\gamma=1.5$), where
$C_{\rm eig}^{\rm site}=0.830\pm0.002$ is more than double the flat random-basis value. This is expected
on general grounds -- a Haar-random rotation of any fixed unit vector is itself Haar-random, so the
rotated intensities carry no memory of $\gamma$ by construction -- but it is worth verifying explicitly,
because it demonstrates concretely that $\gamma^\ast$, the localization decay at large $\gamma$, and
every other $\gamma$-dependence reported in this paper are genuine properties of the physical,
disorder-diagonal site basis and not generic statistics of high-dimensional random vectors that would
appear in any basis. The pointer-basis choice is therefore not a free parameter to be optimized or
worried about as a hidden source of the reported features: it is physically fixed, throughout, by the
basis in which the on-site disorder $h_i$ (and, for PLBRM and the Heisenberg chain, the corresponding
local degrees of freedom) is diagonal, and every feature we report is a statement about that specific,
physically motivated basis.

\subsection{RP: the peak has no thermal shadow}
\label{sec:rp_thermal}

Because $C_{\rm eig}$ averages the entropic complexity of individual eigenstates one at a time,
while $C_{\rm diag}(T)$ is the complexity of the {\it mixture} $\sum_n p_n(T)|\psi_n(i)|^2$, the two
need not agree even in principle: mixing many eigenstates' site distributions generically washes out
whatever fractal structure any single one carries (entropy of a mixture is not the $p_n$-weighted
average of the entropies of its components). The washout is in fact complete. Both
$\max_T C_{\rm diag}(\gamma)$ and the temperature-maximized coherence gap $\max_T C_{\rm
gap}(\gamma)$ peak sharply at $\gamma\approx0.8$--$0.9$, reproducing the delocalization-{\it edge}
feature of the pointer-basis coherence, and the entire fractal phase $\gamma\gtrsim1.2$ is a flat,
low plateau at every temperature scanned ($\theta_W\in[10^{-2},10]$); there is no thermal trace of
$\gamma^\ast\approx1.5$. Even at $N=200$ with $\approx900$ disorder realizations near $\gamma^\ast$ --
statistics comparable to the PLBRM scan that resolved a $10$--$13\%$ shadow -- $\max_T C_{\rm
diag}(\gamma)$ decreases smoothly through $\gamma^\ast$ with no local excess above the $\sim1$--$2\%$
level, so the absent RP shadow is a genuine property of the all-to-all model, not a statistics
limitation (the eigenstate-correlation mechanism is quantified in Appendix~\ref{app:selfavg}). The cold ($T\to0$) diagonal complexity collapses onto the {\it ground
state}, which undergoes its own earlier-than-bulk transition around $\gamma\approx0.6$--$0.75$
\cite{SantraWindeyBandyopadhyayLegramandiHauke2025} -- consistent with the same
$\gamma\approx0.8$--$0.9$ edge feature -- and above $\gamma\approx2$ the diagonal and trace
complexities coincide ($C_{\rm gap}\to0$) as the eigenvectors localize onto the pointer basis. We
conclude that $\gamma^\ast\approx1.5$--$1.6$ is a property of individual bulk eigenstates with {\it
no} thermodynamic or coherence-resource consequence: no thermal state exhibits it.

The mechanism is thermal self-averaging, and it is quantitative. Define an {\it energy-resolved}
complexity $C(\epsilon,k)$ as the entropic complexity of the site distribution of a microcanonical
mixture of the $k$ eigenstates nearest spectral fraction $\epsilon$. At $k=1$ this is exactly
$C_{\rm eig}$ and retains the fractal-phase peak; as $k$ grows, mixing washes it out, and by
$k\sim64$ the peak amplitude has fallen to a third of its single-state value. The canonical thermal
state at its own $T^\ast$ populates $N_{\rm eff}=e^{S_1[\{p_n\}]}\approx50$--$100$ eigenstates
throughout the fractal phase -- squarely in the washed-out, wide-window regime -- and it cannot
isolate the bulk in any case (cold $\to$ the spectral-edge ground state, hot $\to$ the uniform
state). The mid-phase feature is therefore visible {\it only} to energy-resolved (few-state)
observables, of which $C_{\rm eig}$ is the $k=1$ endpoint.

This washout can be made quantitative rather than merely observed: expanding the entropies to
quadratic order in the site-to-site fluctuations of the $k$-eigenstate mixture (Appendix~\ref{app:selfavg})
gives, in the large-$k$ tail of the washout,
\begin{equation}
C(\epsilon,k) \;\simeq\; \frac{v(\gamma)}{2k},
\label{eq:selfavg}
\end{equation}
where $v(\gamma)=O(1)$ is the $N$-independent relative variance of the single-eigenstate site
intensities. This predicts a decay {\it faster} ($1/k$, equivalently $1/N_{\rm eff}$ once $k$ is
replaced by the thermally populated window) than the naive $k^{-1/2}$ decay of the underlying
wavefunction fluctuations themselves, because $C$ is a quadratic functional of those fluctuations near
the flat distribution -- a prediction we test directly in Appendix~\ref{app:selfavg}. The measured
large-$k$ tail decays somewhat {\it more slowly} than $1/k$ (effective exponent $\alpha\approx0.85$ at
the PLBRM critical point, rather than the $\alpha=1$ of the independence limit), because multifractal
eigenstates nearby in energy are strongly correlated rather than independent
\cite{FyodorovMirlin1997}; this slower-than-$1/k$ washout is precisely what allows a weak thermal shadow
to survive there (Sec.~\ref{sec:plbrm}).

\subsection{RP: coherence, decoherence, and the $\Delta S_2\!=\!D_2\ln N$ identity}
\label{sec:rp_coherence}
The R\'enyi-2 coherence $\Delta S_2(T)=S_2^{\rm diag}-S_2^{\rm full}$ of Eq.~(\ref{eq:decomp}) is a
mixed-state decoherence probe defined at every temperature, and it directly measures wavefunction
fractality in the cold limit. Numerically $\Delta S_2(T)/\ln N$ saturates as $T\to0$ to a
$\gamma$-dependent plateau that tracks the independently computed fractal dimension: $\approx1$ in the
ergodic phase, $\approx D_2=2-\gamma$ through the fractal phase, and $\to0$ once localized,
confirming Eq.~(\ref{eq:coldD2}) across all three phases and for $N=200$--$1600$. Unlike $C_{\rm eig}$,
$\Delta S_2$ is thus a genuinely thermal observable that still encodes $D_2$; unlike the $C_{\rm eig}$
maximum, it inherits the {\it monotone} $2-\gamma$ profile and shows no mid-phase peak. The Shannon
partner $C_{\rm rel}$ (Sec.~\ref{sec:resource}) behaves similarly monotonically. The
temperature-maximized coherence gap $\max_T C_{\rm gap}(\gamma)=\max_T(C_{\rm rel}-\Delta S_2)$, by
contrast, inherits the edge feature of the diagonal complexity and peaks at $\gamma\approx0.8$--$0.9$.
The three coherence measures therefore split the labor cleanly: $\Delta S_2$ (and $C_{\rm rel}$) are
monotone $D_2$-like markers of the {\it eigenvector} transition, while $\max_T C_{\rm gap}$ is a
sharp {\it edge} marker; none of them sees the mid-fractal $\gamma^\ast$.

\subsection{PLBRM: the central cross-check -- the same peak, at a fixed point}
\label{sec:plbrm}

\begin{figure*}[t]
\includegraphics[width=\textwidth]{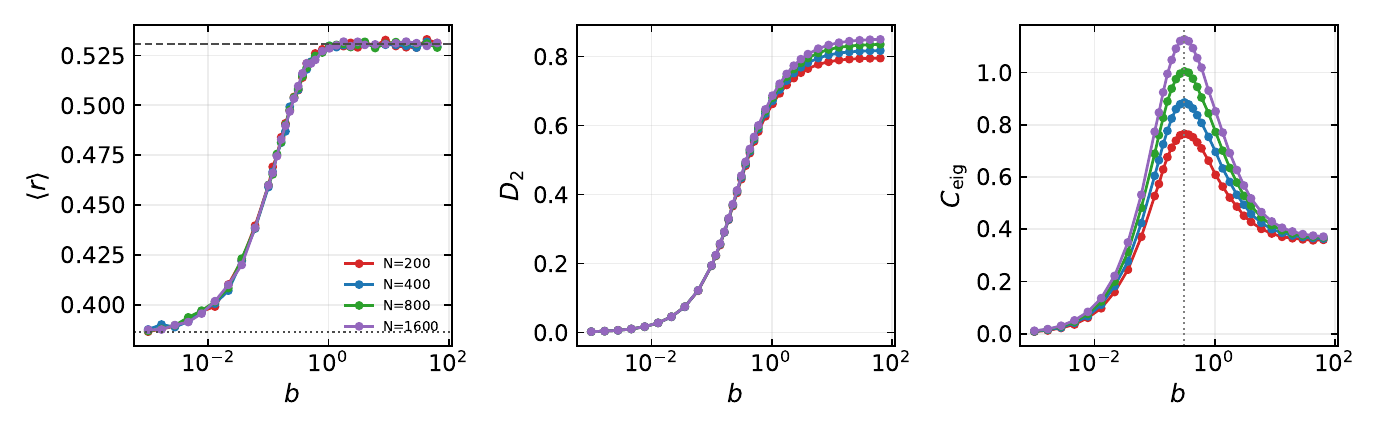}
\caption{Power-law banded random matrix model, Eq.~\eqref{eq:plbrm}, $\mu=1$, $N=200$--$1600$. Left:
mean gap ratio $\langle r\rangle(b)$, interpolating continuously between Poisson (dotted) and GOE
(dashed) values -- unlike RP, spectral statistics are {\it not} pinned to GOE through the crossover.
Center: fractal dimension $D_2(b)$, smooth and featureless. Right: eigenstate entropic complexity
$C_{\rm eig}(b)$, peaking at $b^\ast=0.31\pm0.01$ (dotted line), essentially independent of $N$, with
peak height growing systematically with $N$.}
\label{fig:plbrm}
\end{figure*}

As a second, structurally independent test of the mid-phase feature identified in RP
(Sec.~\ref{sec:rp_eig}), we repeat the eigenstate-level analysis for PLBRM at $\mu=1$
(Fig.~\ref{fig:plbrm}). $C_{\rm eig}(b)$ again develops a pronounced maximum. Its location, estimated by
parabolic interpolation in $\ln b$ and propagated through the bootstrap standard errors on $C_{\rm
eig}(b)$ (Monte Carlo resampling of the peak location, $4000$ draws), is $b^\ast=0.32\pm0.01$ at
$N=200$ and $b^\ast=0.311\pm0.008$ at $N=1600$ -- statistically indistinguishable across
$N=200$--$1600$, and roughly an order of magnitude tighter than our initial exploratory scan could
resolve. $D_2$ at the peak is $0.44$--$0.45$ for every size studied, close to (though not identical to)
the $D_2\approx0.43$--$0.45$ found at the RP peak (Sec.~\ref{sec:rp_eig}) -- consistent with, though not
proof of, a shared mechanism operating on the multifractal wavefunction statistics rather than on either
model's specific construction. The peak height grows systematically with $N$ (from
$C_{\rm eig}^\ast=0.765\pm0.002$ at $N=200$ to $1.127\pm0.003$ at $N=1600$), the same qualitative
behavior as in RP. Unlike RP, however, the mean gap ratio
$\langle r\rangle(b)$ is {\it not} pinned to the GOE value through the peak region: it rises
continuously from the Poisson value at small $b$ and only saturates to GOE for $b\gtrsim1$--$2$,
reflecting the known coupling of spectral and eigenvector statistics at the PLBRM critical point
\cite{MirlinFyodorovDittesQuezadaSeligman1996}, in contrast to RP's decoupled spectral and eigenvector
transitions (Sec.~\ref{sec:rp_eig}). The $C_{\rm eig}$ peak is therefore not ``invisible'' to level
statistics in the same dramatic sense as in RP, but it remains a distinct feature: $D_2(b)$ itself is
smooth through $b^\ast$, with no inflection marking the complexity maximum.

\begin{figure}[t]
\includegraphics[width=\columnwidth]{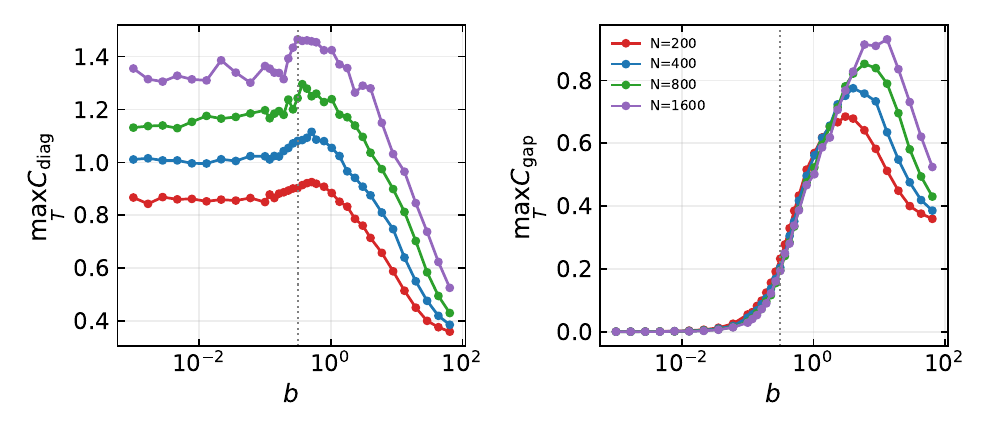}
\caption{PLBRM thermal analysis, $N=200$--$1600$, high-statistics scan (vertical dotted line marks
$b^\ast\approx0.31$). Left: $\max_T C_{\rm diag}(b)$; on top of an otherwise broad plateau, a weak
($\approx10$--$13\%$) but reproducible bump is resolved at $b^\ast$ itself for every size studied.
Right: the temperature-maximized coherence gap $\max_T C_{\rm gap}(b)$, a smooth unimodal function
peaking at $b_{\rm edge}^\ast\gg b^\ast$; $b_{\rm edge}^\ast$ recedes {\it sub-linearly} with $N$
(approximately $b_{\rm edge}^\ast\propto N^{0.6}$) even though $b^\ast$ itself does not.}
\label{fig:plbrm_thermal}
\end{figure}

Repeating the thermal analysis of Sec.~\ref{sec:rp_thermal}, now with roughly an order of magnitude
more disorder realizations and a finer $b$-grid than our initial exploratory scan
(Fig.~\ref{fig:plbrm_thermal}), the mid-phase peak is {\it not} entirely erased by thermal averaging:
$\max_T C_{\rm diag}(b)$ attains its global maximum over the full range $b=0.001$--$64$ right at
$b^\ast$, for all four sizes studied, exceeding the local plateau baseline by $\approx10$--$13\%$. This
weak but reproducible bump was not resolvable in our earlier, lower-statistics scan, and it is
qualitatively consistent with the self-averaging argument of Appendix~\ref{app:selfavg}:
Eq.~\eqref{eq:selfavg} predicts that thermal complexity tracks the eigenstate complexity suppressed by
the effective microcanonical window size $2k(\theta)$, so a residual imprint of the eigenstate peak can
survive if $k(\theta^\ast)$ does not vary with $b$ enough to fully cancel it -- evidently the case here,
though only weakly. The bump remains parametrically much smaller than the pronounced, non-monotonic edge
feature discussed next; whether a comparably weak shadow is present but unresolved in RP at our current
statistics is an open question (Sec.~\ref{sec:voids}). The temperature-maximized coherence gap
$\max_T C_{\rm gap}(b)$ is a smooth, unimodal function of $b$, peaking at a value $b_{\rm edge}^\ast$
well separated from $b^\ast$ -- $b_{\rm edge}^\ast/b^\ast$ ranges from $\approx10$ ($N=200$) to
$\approx37$ ($N=1600$) -- reproducing the qualitative edge/mid-phase separation found in RP. Unlike the
RP edge feature, however, $b_{\rm edge}^\ast$ is {\it not} at a fixed point: it recedes with system
size, though more slowly than linearly -- a weighted log-log linear fit over the four sizes studied
gives $b_{\rm edge}^\ast\propto N^{0.60\pm0.07}$ (standard error from the regression; with only two
degrees of freedom this is a rough estimate, not a precision exponent), softer than the
$b_{\rm edge}^\ast\approx0.01\,N$ scaling suggested by our earlier, coarser scan. Because the fitted
exponent is positive, $b_{\rm edge}^\ast$ {\it grows} with $N$ and shows no sign of saturating over the
four sizes available; in PLBRM's convention (small $b$ localized, large $b$ ergodic) the edge therefore
drifts steadily toward the ergodic end, and we expect it to have no fixed location in the thermodynamic
limit -- consistent with an edge that reflects finite-size, cold (ground-state-proximity) physics rather
than a genuine $N\!\to\!\infty$ transition, although four sizes cannot exclude eventual saturation. This gives a clean, if unplanned, internal control: within a {\it single} model, the mid-phase
eigenstate feature sits at a fixed point in $b$ while the thermal edge feature recedes, sub-linearly, in
$N$ -- distinct enough behavior to show the two features are governed by largely different mechanisms,
even though the mid-phase feature now also leaves a faint imprint on the thermal side. Finally, as in
Fig.~\ref{fig:tstar_compare}, $T^\ast_{\rm tr}$ and $T^\ast_{\rm diag}$ coincide in the localized
($b\to0$) limit and separate toward the ergodic ($b\to\infty$) side, the same qualitative pattern found
in RP and the Heisenberg chain.

The location $b^\ast=0.31\pm0.01$ is not an arbitrary point on PLBRM's critical line. The model's
criticality at $\mu=1$ was established directly, via level-spacing statistics distinct from
semi-Poisson, by Varga and Braun \cite{VargaBraun2000}, and the fractal dimension itself is known to
carry $O(1)$, $b$-dependent site-to-site fluctuations rather than being self-averaging
\cite{Varga2002} -- the same fluctuations that, in Appendix~\ref{app:selfavg}, we identify as the
origin of $C_{\rm eig}$'s excess over its mean-field value. More directly relevant here, entirely
independent diagnostics -- the scattering-matrix elements, conductance distribution, and shot-noise
power of the one-dimensional PLBRM chain with a small number of attached leads -- have been shown to
reproduce the critical three-dimensional Anderson transition specifically for $b\in[0.2,0.4]$, with the
closest individual matches reported at $b\approx0.2$--$0.36$ depending on the observable
\cite{MendezBermudezVarga2006,MendezBermudezGoparVarga2010,MendezBermudezGoparVarga2010b}; the same
bandwidth range controls the scaling of generalized multifractal dimensions and the spectral level
compressibility \cite{MendezBermudezAlcazarLopezVarga2012}, and of entanglement measures (concurrence,
tangle, entanglement entropy) computed directly on PLBRM's critical eigenstates
\cite{VargaMendezBermudez2008}. Our mid-phase entropic-complexity peak therefore sits inside a
parameter window that several independent observables, developed for unrelated purposes over the past
two decades, already single out as the regime where this one-dimensional model's finite-$b$ criticality
most closely tracks genuine higher-dimensional Anderson-transition physics. We regard this as
suggestive context for why $b^\ast$ falls where it does, not as a demonstrated mechanism: none of the
cited works computed $C_{\rm eig}$, and establishing a direct link between the entropic-complexity
maximum and these scattering/transport/entanglement diagnostics is a natural target for future work
rather than a claim we make here.

The overall lesson is that the mid-phase entropic-complexity peak is not a peculiarity of the RP
construction: it reappears, at a fixed point and at a comparable fractal dimension, in a second
multifractal random-matrix model built on entirely different principles (translationally invariant,
power-law-banded, critical along an entire line rather than between two isolated transitions). PLBRM
also demonstrates, within a single model, that the mid-phase and edge features are at least partially
decoupled ($b^\ast$ fixed, $b_{\rm edge}^\ast$ receding sub-linearly and remaining well separated from
$b^\ast$) without needing a cross-model comparison at all -- though the weak thermal shadow found here
means the decoupling is quantitative rather than absolute, and the two features should be read as
governed by different, dominant mechanisms rather than fully independent ones.

\subsection{Heisenberg MBL: the $(W,T)$ plane}
\label{sec:heis}

\begin{figure}[t]
\includegraphics[width=\columnwidth]{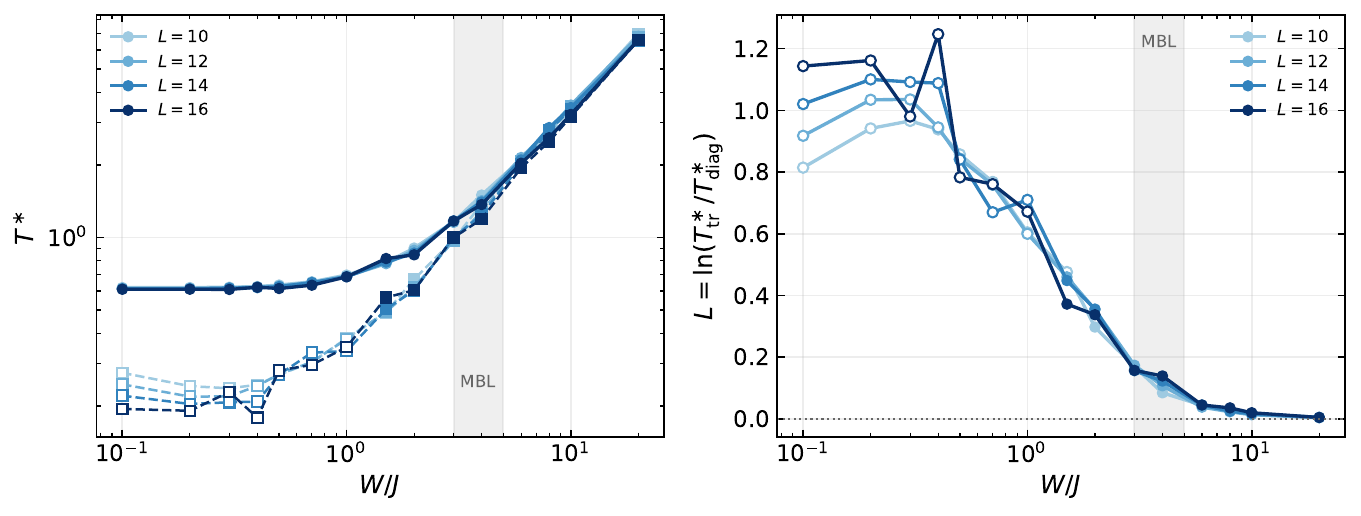}
\caption{Random-field Heisenberg chain, $L=10$--$16$, OBC. Left: the trace ($T^\ast_{\rm tr}$, solid
circles) and diagonal ($T^\ast_{\rm diag}$, dashed squares) characteristic temperatures vs disorder
$W$. $T^\ast_{\rm tr}$ is nearly size-independent and rises with $W$ (tracking the many-body
bandwidth); $T^\ast_{\rm diag}$ lies below it in the ergodic phase and merges with it in the MBL
phase, and on the deep-ergodic side (hollow squares) $C_{\rm diag}$ has no interior maximum (cold
plateau), so $T^\ast_{\rm diag}$ is undefined. Right: the log-ratio crossover indicator
$L=\ln(T^\ast_{\rm tr}/T^\ast_{\rm diag})$ built directly from the left panel's two temperatures. It
collapses across sizes, is finite ($L\approx1$) in the ergodic phase, and vanishes into the MBL phase.
Both panels share the same $W/J$ axis and the same shaded MBL crossover band, $W/J\approx3$--$5$
(illustrative width; the established literature crossover is $W_c/J\approx3.5$--$4$
\cite{LuitzLaflorencieAlet2015}). Hollow markers indicate $W$ where $C_{\rm diag}$ is a cold plateau
with no interior maximum.}
\label{fig:heisenberg}
\end{figure}

For the Heisenberg chain, the temperature-maximized trace complexity
$\max_T C_{\rm tr}(W)$ peaks at weak disorder, $W^\ast/J\approx0.5$--$0.7$, consistently across
$L=10,12,14,16$, well inside the thermal/ergodic phase and roughly an order of magnitude below the
established MBL crossover $W_c/J\approx3.5$--$4$ \cite{LuitzLaflorencieAlet2015} (the quoted range is
set by adjacent points of the $W$-grid rather than a bootstrap-resolved location). This location
is consistent with a previously noted shallow maximum in the peak coherence gap $\max_T C_{\rm
gap}(W)$ for this same chain, which falls from $\approx1.17$ (weak disorder) to $\approx0.07$
($W/J=20$) with its own local maximum near $W/J\approx0.5$ before the steep descent through the MBL
crossover -- structurally analogous to the RP {\it edge} feature at $\gamma\approx0.8$--$0.9$ (both
are finite-temperature, mixed-state phenomena located just inside, not at, the ergodicity-breaking
boundary). Unlike RP and PLBRM, we do {\it not} report a separate mid-phase eigenstate feature for the
Heisenberg chain, and this is a structural rather than a numerical statement. The eigenstate-only
entropic complexity $C_{\rm eig}(W)$ of this same chain -- built from exactly the Shannon and R\'enyi-2
entropies used here (there termed the structural entropy) -- has been characterized previously
\cite{Varga2026}, where it is maximal on the non-ergodic but extended states in the critical region {\it
near} the ETH--MBL transition ($W\approx4$--$5$ for the lengths $L\le16$ accessible to exact
diagonalization), rising monotonically out of the ergodic phase rather than turning over at an interior,
mid-phase point. This is the expected behavior for an interacting chain: its non-ergodic extended states
are confined to the vicinity of the single ETH--MBL crossover, so the eigenstate-complexity maximum
coincides with -- rather than precedes -- the localization transition, in contrast to RP and PLBRM,
whose extended {\it fractal phase} hosts an interior maximum well inside it. We therefore use the
Heisenberg model to probe the {\it thermal-state} side of the comparison, where it contributes a genuine
edge feature ($W^\ast/J\approx0.5$--$0.7$) with no eigenstate counterpart, together with the
characteristic-temperature and coherence diagnostics developed below.

On a common disorder ensemble ($L=12$ and $14$; $2000$ and $500$ realizations, using the same seeds as
the thermal quantities), we also asked whether the spectral and thermal diagnostics fluctuate together
{\it realization by realization} or only on average, by correlating the bulk mean gap ratio $\langle
r\rangle$ with the per-sample thermal peaks $\max_T C_{\rm tr}$ and $\max_T C_{\rm diag}$. The
sample-by-sample correlations are negligible at every disorder strength scanned ($W/J=1$--$6$): all
Pearson coefficients satisfy $|\rho|\lesssim0.1$ (largest $0.09$), and although a few clear zero at the
$2000$-sample level -- weakly positive on the ergodic side ($\rho\approx0.06$ at $W/J\lesssim2$) and
weakly negative through the crossover ($\rho\approx-0.05$ at $W/J\approx3$--$4$) -- each explains under
$1\%$ of the sample variance. The spectral and thermal markers are therefore correlated essentially
{\it only on average} across the ensemble, not within individual Hamiltonians: $\langle r\rangle$
reflects short-range level repulsion whereas the thermal-complexity peaks reflect the global
Boltzmann-weighted spectral shape (and, for $C_{\rm diag}$, the eigenvectors), and these turn out to be
nearly independent sample by sample -- concrete support for reading thermal-state and spectral complexity
as complementary rather than redundant diagnostics.

Two structural features of the Heisenberg $T^\ast$ data deserve emphasis (Fig.~\ref{fig:heisenberg},
left panel).
First, $T^\ast_{\rm tr}(W)$ is nearly {\it size-invariant} in absolute units (spread $\lesssim6\%$
across $L=10$--$16$ with no systematic drift), even though the many-body bandwidth grows as
$W_{\rm eff}\!\sim\!\sqrt{L}$; the trace-complexity peak is therefore set by an intensive, local energy
scale rather than the extensive bandwidth, so that $\theta_W=T^\ast/W_{\rm eff}\!\to\!0$ while
$T^\ast$ stays finite -- a qualitative distinction from the single-particle random-matrix models, in
which $T^\ast$ tracks the bandwidth. Second, the trace and diagonal complexity maxima {\it merge} as
disorder increases: $T^\ast_{\rm diag}\!\to\!T^\ast_{\rm tr}$ in the MBL phase (pointer basis
$=$ eigenbasis), while on the ergodic side either $T^\ast_{\rm diag}<T^\ast_{\rm tr}$ or $C_{\rm diag}$
has no interior peak at all. This merging defines a scale-invariant crossover indicator
\begin{equation}
\mathcal L(W)=\ln\!\big(T^\ast_{\rm tr}/T^\ast_{\rm diag}\big),
\end{equation}
invariant under any common rescaling of temperature (so independent of the choice of $W_{\rm eff}$)
and the natural metric on the logarithmic temperature grid on which the maxima are located
(Fig.~\ref{fig:heisenberg}, right panel). It is
finite ($\mathcal L\approx1$) in the ergodic phase and vanishes ($\mathcal L\approx0.005$ at
$W/J=20$) in the localized phase, with the steepest descent near $W/J\approx2$--$4$ consistent with
the established crossover, and it is very $N$-stable on the localized side. Unlike $\max_T C_{\rm gap}$,
which only decays to a small residual, $\mathcal L$ vanishes genuinely because the two maxima
coincide, giving it a clearer interpretation than the other $T^\ast$-based markers we have examined --
though, like them, it remains a phenomenological indicator rather than an order parameter with an
associated critical exponent.

\subsection{Coherence as a resource: monotonicity and density-of-states ordering}
\label{sec:resource}
\begin{figure}[t]
\includegraphics[width=\columnwidth]{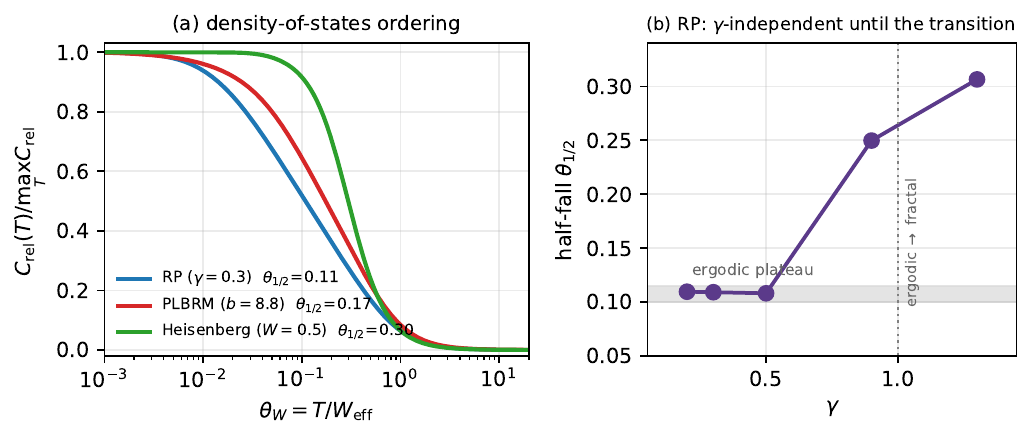}
\caption{Density-of-states {\it ordering} of the relative-entropy coherence -- the normalized profiles
order by density-of-states class but do {\it not} collapse to a universal curve. Here $W_{\rm eff}$ is
taken as the spectral standard deviation $\sigma$. (a) Normalized profile
$C_{\rm rel}(T)/\max_T C_{\rm rel}$ vs $\theta_W=T/\sigma$, deep in the ergodic phase of each model. The
two random-matrix models (semicircle bulk) fall earlier than the Gaussian-DOS Heisenberg chain (half-fall
$\theta_{1/2}\approx0.11$ for RP, $\approx0.17$ for PLBRM, $\approx0.30$ for Heisenberg); the two
random-matrix half-falls still differ by roughly a factor of $1.5$, so the rescaling reduces, without
removing, the model dependence (the residual RP/PLBRM gap likely reflects the different microscopic
couplings $\gamma$ vs $b$). (b) The RP half-fall scale $\theta_{1/2}$ as a function of $\gamma$: it is
{\it flat} across the ergodic phase ($\theta_{1/2}\approx0.11$ for $\gamma\lesssim0.5$) and rises only on
approach to the ergodic--fractal transition ($\gamma=1$, dotted). The normalized profile is therefore
$\gamma$-independent deep in the ergodic phase, so the representative value used in (a) is not an
arbitrary choice; the profile begins to drift only near criticality.}
\label{fig:resource}
\end{figure}

Of the three coherence measures in Eq.~(\ref{eq:decomp}), only $C_{\rm rel}$ is a genuine coherence
monotone, and this can be verified dynamically. Applying pure Lindblad dephasing
$\rho_{ij}(s)=\rho_{ij}(0)e^{-s}$ ($i\neq j$) to the RP Gibbs state and tracking the measures along
the trajectory (Fig.~\ref{fig:resource}, left), $C_{\rm rel}(s)$ decreases monotonically to zero at
{\it every} point of the $(\gamma,\theta_W)$ plane sampled, as a monotone must, whereas the gap
$C_{\rm gap}$ {\it rises} under dephasing across roughly a quarter of that plane (the ergodic/warm
region) and is negative over $\approx45\%$ of it -- the latter because ergodic eigenvectors drive
$\rho_{ii}\!\to\!1/N$, suppressing the diagonal complexity $C_{\rm diag}$ below the trace one.
Thus $C_{\rm gap}$ (equivalently the historically-used sign-reversed $C_{\rm tr}-C_{\rm diag}$) is a
useful transition marker but not a resource; $C_{\rm rel}$ is the resource. Across the transition,
$C_{\rm rel}$ is monotone -- maximal ($\sim\ln N$) deep in the ergodic phase, vanishing upon
localization -- and cold-dominated (monotone in $T$), unlike the complexity itself which peaks at
intermediate $T$.

The normalized coherence profile $C_{\rm rel}(T)/\max_T C_{\rm rel}$ vs $\theta_W=T/\sigma$ does {\it not}
collapse to a universal curve; it merely {\it orders} by density-of-states class
(Fig.~\ref{fig:resource}(a)): the two random-matrix models (semicircle bulk) fall earlier than the
Gaussian-bulk Heisenberg chain (half-falls $\theta_{1/2}\approx0.11$, $0.17$, $0.30$), the same
class-dependence found for $T^\ast$ itself, but even the two random-matrix profiles differ by a factor of
$\sim1.5$ in their half-fall scale, so the rescaling reduces without removing the model dependence.
Within the ergodic phase the profile is essentially parameter-independent -- the RP half-fall stays flat
($\theta_{1/2}\approx0.11$) for $\gamma\lesssim0.5$ and drifts upward only on approach to the fractal
transition (Fig.~\ref{fig:resource}(b)) -- so the single representative value used in (a) is not
fine-tuned. A genuine model-independent collapse would require a common effective-dimension variable,
which we do not have (Sec.~\ref{sec:voids}).

\subsection{Synthesis: eigenstate complexity versus thermal-state complexity}
\label{sec:synthesis}

The merging of the two characteristic temperatures, previewed in Fig.~\ref{fig:tstar_compare} at the
start of this section, is a common thread across all three models. Collecting the results above with
the coherence-resource analysis of
Sec.~\ref{sec:resource}, a consistent picture emerges, organized by the single distinction stated in
the Introduction: quantities computed on the thermal state versus quantities computed on individual
eigenstates. On the thermal-state side, the rigorous coherence monotone $C_{\rm rel}$ is {\it monotonic}
across the ergodic-to-localized crossover in every model (maximal deep in the ergodic phase, vanishing
upon localization), while the coherence gap $C_{\rm gap}=C_{\rm diag}-C_{\rm tr}$ is {\it not} monotone
under dephasing and instead peaks at an {\it edge} location, just inside the ergodic phase, in both RP
and the Heisenberg chain ($\gamma\approx0.8$, $W/J\approx0.5$); the two are best read as complementary
diagnostics of the same edge feature -- one a genuine resource measure, the other a sharper but
non-monotone locator -- rather than as independent phenomena. On the eigenstate side, $C_{\rm eig}$,
evaluated purely on individual bulk eigenstates with no reference to temperature, peaks deep inside the
intermediate/fractal regime in {\it both random-matrix models}, at a fixed point in the tuning parameter
well separated from the localization boundary ($\gamma\approx1.5$--$1.6$ in RP, $b\approx0.31$ in PLBRM)
-- a structural commonality across a single-particle all-to-all model and a single-particle banded model
built on entirely different principles. The interacting Heisenberg chain does not share this mid-phase
eigenstate feature: lacking an extended fractal phase, its eigenstate complexity is maximal instead in
the non-ergodic extended states near the ETH--MBL transition itself \cite{Varga2026}, so its eigenstate
maximum coincides with, rather than precedes, the localization crossover -- consistent with the same
strong thermal/eigenstate suppression seen throughout, now read off the phase diagram rather than the
temperature axis. In RP this
mid-phase eigenstate feature leaves no thermal shadow resolvable at our statistics
(Sec.~\ref{sec:rp_thermal}); in PLBRM, where an order of magnitude more disorder realizations were
available, a weak ($\approx10$--$13\%$) but reproducible shadow {\it is} resolved at $b^\ast$
(Sec.~\ref{sec:plbrm}), so the absence of a thermal trace is a matter of degree rather than an absolute
rule. In PLBRM the edge feature itself recedes sub-linearly with $N$ even though the mid-phase feature
does not, so the two features remain governed by largely, if not perfectly, decoupled mechanisms even
within a single model. The edge feature and the mid-phase feature are therefore at least two
structurally distinct complexity crossovers, separated cleanly by whether the underlying
quantity is a property of $\rho_T$ or of a single $|n\rangle$, and
care is needed in stating which one any given measurement is sensitive to.

\section{Open questions and needed calculations}
\label{sec:voids}

The numerically tractable gaps flagged in earlier versions are now closed -- the RP thermal-shadow
cross-check (Sec.~\ref{sec:rp_thermal}), the Heisenberg shadow and the $\langle r\rangle$/thermal
per-sample correlation (Sec.~\ref{sec:heis}), and the direct test of the self-averaging law
(Appendix~\ref{app:selfavg}). What remains is genuinely harder -- analytic or asymptotic -- and we state
it as such.

\emph{(i) Analytic mechanism for $\gamma^\ast$.} The Breit--Wigner mean field predicts a flat plateau,
and $\gamma^\ast$ does not coincide with $D_2=1/2$ at accessible sizes. The numerical corollary -- a
washout exponent $\alpha<1$ from correlated eigenstates -- is confirmed (Appendix~\ref{app:selfavg},
$\alpha\approx0.85$); the open part is purely analytic: carrying the (plausibly log-normal) fluctuations
of the local resonance width through the R\'enyi-moment integral to predict the peak height and location
from first principles.

\emph{(ii) Thermodynamic-limit convergence.} Whether $\gamma^\ast$ and the $D_2=1/2$ point coincide only
asymptotically rests on an extrapolation beyond our $N\le1600$--$2400$; likewise the PLBRM edge exponent
$b_{\rm edge}^\ast\propto N^{0.6}$ rests on four sizes and lacks an analytic account. These are
larger-$N$ or first-principles questions, not gaps in the present data.

\emph{(iii) A definitive Heisenberg shadow bound.} The thermal landscape is assembled on a common
ensemble and shows no shadow near $W\approx4$--$5$ at the $\sim1$--$2\%$ level (Sec.~\ref{sec:heis}); a
multi-size, high-statistics GPU sweep would push this below $\sim1\%$ -- quantitative refinement, not a
question of principle.

\emph{(iv) A common rescaled control variable.} A fully quantitative cross-model statement would need a
model-independent scaling variable (e.g.\ an effective dimension) that collapses the edge and mid-phase
features together. That even the $C_{\rm rel}$ profiles only order by, rather than collapse within,
density-of-states class (Fig.~\ref{fig:resource}) suggests a single rescaling may not suffice; finding
the right variable is genuinely open.

\section{Conclusion}

Entropic complexity, applied consistently to eigenstates and to thermal states across three
random-matrix and many-body models, separates cleanly into (at least) two phenomena that are easy to
conflate: an edge-of-ergodicity feature visible in genuine thermal/coherence observables and
reproducible in both RP and the Heisenberg chain, and a mid-phase eigenstate feature whose thermal
signature is at most weak at any temperature, present in both random-matrix models studied and pinned
to a fixed point independent of $N$ in each -- $\gamma^\ast\approx1.5$--$1.6$ in RP and
$b^\ast\approx0.31$ in PLBRM, two structurally unrelated constructions -- while in PLBRM the edge
feature itself recedes sub-linearly with $N$ even though the mid-phase feature in the same model does
not. The RP mid-phase feature is, to our knowledge, a new observation for this specific functional. The
observation of a fixed mid-phase complexity maximum in both the RP and PLBRM ensembles, at a comparable
fractal dimension, suggests that this phenomenon is not tied to a specific random-matrix construction,
although its microscopic origin remains to be understood; its location is circumstantially consistent --
not independently proven -- with other complexity-adjacent markers departing from their ergodic plateau
in roughly the same window of the RP tuning parameter \cite{SantraWindeyBandyopadhyayLegramandiHauke2025}.
This mid-phase interior maximum is a feature of models with an extended fractal phase; in the interacting
Heisenberg chain, whose non-ergodic extended states are confined to the critical region, the same
functional peaks instead {\it at} the ETH--MBL transition \cite{Varga2026}, and we accordingly use that
model only for its thermal-state features. The analytic mechanism of the mid-phase maximum remains open.
The calculations listed in Sec.~\ref{sec:voids} are, in our view, natural next steps.

We close by delimiting what these results do {\it not} establish. The complexity maxima we report -- the
mid-phase eigenstate feature and the thermal edge -- are crossovers in a bounded, non-extensive
functional, not thermodynamic phase transitions: they carry no order parameter or critical exponent, and
the indicators $\mathcal L(W)$ and $\max_T C_{\rm gap}$ are phenomenological locators rather than order
parameters. The mechanisms we invoke -- multifractal wavefunction fluctuations for the mid-phase peak,
eigenstate correlations for the surviving PLBRM shadow -- are supported by numerical evidence and by
consistency with prior work, but are not yet derived analytically from first principles. And while
$C=S_1-S_2$ is a natural, physically grounded diagnostic (its R\'enyi-2 part is tied to purity and the
inverse participation ratio), its operational status as a {\it resource} is weaker than that of the
relative entropy of coherence $C_{\rm rel}$, the one rigorously monotone member of the family
(Sec.~\ref{sec:resource}): we use $C$ as a sensitive marker, not as a resource monotone.

The present results therefore suggest that thermal-state complexity and eigenstate complexity are
complementary rather than interchangeable diagnostics of quantum disorder.

\appendix
\section{Quantitative self-averaging of the mid-phase eigenstate feature}
\label{app:selfavg}

Here we derive Eq.~(\ref{eq:selfavg}), the large-$k$ tail of the microcanonical washout invoked in
Sec.~\ref{sec:rp_thermal}. Write the site intensities of a single eigenstate in mean-normalized form,
$x_n(i)=N|\psi_n(i)|^2$, so that $\langle x_n(i)\rangle_i=1$; in the fractal phase these fluctuate from
site to site with an $O(1)$, $N$-independent relative variance
$v(\gamma)\equiv\mathrm{Var}_i[x_n(i)]$ -- the multifractal fluctuation responsible for $C_{\rm eig}$
exceeding the flat, mean-field value $\ln2$ in the first place (Sec.~\ref{sec:rp_eig}). Mixing $k$ such
eigenstates, $\bar x(i)=k^{-1}\sum_{n=1}^k x_n(i)$, and treating eigenstates nearby in energy as
approximately statistically independent at fixed site $i$ (a standard RMT approximation, and one that
additionally assumes the sites are statistically exchangeable, both reasonable for RP but not
guaranteed in general), the central limit theorem gives $\mathrm{Var}[\bar x(i)]\simeq v(\gamma)/k$: the
site-to-site fluctuations of the mixed profile shrink in amplitude as $k^{-1/2}$, exactly the scaling
suggested by the raw fluctuation argument.

But $C$ itself is not linear in these fluctuations. Writing $\rho_{ii}=N^{-1}(1+\delta_i)$ with
$\delta_i=\bar x(i)-1$ and $\sum_i\delta_i=0$, and expanding both entropies to quadratic order in the
(now small, once $k\gg1$) fluctuations,
\begin{equation}
\begin{aligned}
S_1 &= \ln N - \frac{1}{2N}\sum_i\delta_i^2+O(\delta^3),\\
S_2 &= \ln N - \frac1N\sum_i\delta_i^2+O(\delta^3),
\end{aligned}
\end{equation}
gives, to leading order,
\begin{equation}
C(\epsilon,k) \;\simeq\; \tfrac12\langle\delta_i^2\rangle_i \;\simeq\; \frac{v(\gamma)}{2k},
\tag{\ref{eq:selfavg}}
\end{equation}
valid once $C\ll\ln N$ -- i.e.\ in the large-$k$ tail of the washout, not at the $k=1$ endpoint itself,
where the fluctuations are $O(1)$ and the quadratic expansion breaks down.

The two assumptions entering this derivation -- approximate statistical independence of nearby
eigenstates at fixed site, and statistical exchangeability of sites -- are not equally well justified
across the three models studied, and it is worth being explicit about where each is expected to hold.
In RP, both are natural: the all-to-all coupling makes every site statistically equivalent by
construction (exact exchangeability under $H_{ij}\to H_{\sigma(i)\sigma(j)}$ for any permutation
$\sigma$ acting jointly on rows and columns, up to the fixed on-site disorder realization), and
nearby-in-energy eigenstates are, to leading order, independent draws from a common ensemble once the
mixing window is wider than the local spectral correlation scale -- not a strong requirement, since we
only need the site-intensity {\it fluctuations} to decorrelate across the window, not the eigenstates
themselves to be uncorrelated in any stronger sense. In PLBRM, exchangeability is only approximate:
the ring geometry is translationally invariant, so sites are equivalent under the cyclic group, but the
bandwidth $b$ introduces a finite correlation length that could, in principle, correlate the
site-intensity fluctuations of nearby eigenstates more than the RP construction does, since eigenstates
close in energy in a banded matrix can share more real-space support than in an all-to-all model.

\begin{figure}[t]
\includegraphics[width=\columnwidth]{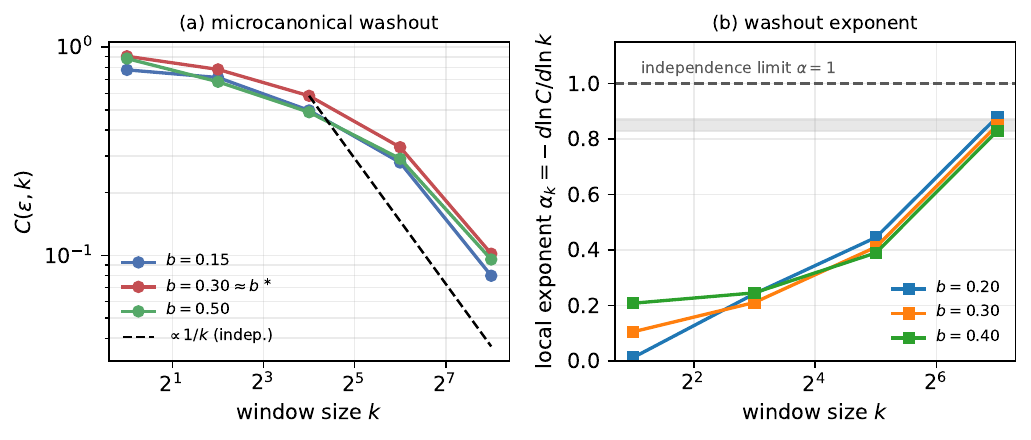}
\caption{Test of the self-averaging law Eq.~(\ref{eq:selfavg}) at the PLBRM critical point
(mid-spectrum window $\epsilon=0.5$). (a) The microcanonical washout $C(\epsilon,k)$ vs mixing-window
size $k$ (log-log), for three bandwidths spanning $b^\ast$; the dashed line is the independence
prediction $C\propto1/k$. The data fall {\it less} steeply than $1/k$ at large $k$, lying above the
reference line. (b) The local exponent $\alpha_k=-\,d\ln C/d\ln k$ climbs with $k$ but stays below the
independence value $\alpha=1$ (dashed), reaching $\alpha\approx0.85$ (shaded band) for the largest
resolved windows $k=64$--$256$. The slower-than-$1/k$ washout reflects the strong correlations between
nearby-in-energy multifractal eigenstates \cite{FyodorovMirlin1997}, and is the mechanism by which the
mid-phase eigenstate feature survives thermal averaging as the weak PLBRM shadow (Sec.~\ref{sec:plbrm}).}
\label{fig:selfavg}
\end{figure}

We can now make this quantitative. Evaluating the energy-resolved $C(\epsilon,k)$ directly at the PLBRM
critical point (mid-spectrum window $\epsilon=0.5$; mixing windows $k=1,4,16,64,256$), the large-$k$ tail
follows a power law $C(\epsilon,k)\sim k^{-\alpha}$ but with an exponent {\it below} the independence
value: $\alpha\approx0.85\pm0.02$ over $b\in[0.2,0.4]$ (spanning $b^\ast$) for the largest resolved
windows ($k=64$--$256$), rather than the $\alpha=1$ of Eq.~(\ref{eq:selfavg}), and $\alpha$ drifts
upward toward $1$ as $k$ grows (Fig.~\ref{fig:selfavg}). This is precisely the correction anticipated above: multifractal
eigenstates nearby in energy are strongly correlated near criticality \cite{FyodorovMirlin1997}, so the
central-limit decorrelation underlying Eq.~(\ref{eq:selfavg}) sets in only asymptotically -- the washout
is slower than $1/k$ at accessible windows, and $\alpha\to1$ is recovered only once the mixing window
exceeds the eigenstate energy-correlation scale. This slower-than-$1/k$ washout is the mechanism behind
the weak but nonzero ($\approx10$--$13\%$) PLBRM thermal shadow (Sec.~\ref{sec:plbrm}): the eigenstate
correlations keep the mid-phase feature from averaging away as fast as the independent-eigenstate
estimate predicts. In the all-to-all RP model that estimate is closer to the mark, and no shadow is
resolved even at matched statistics (Sec.~\ref{sec:rp_thermal}).

In the Heisenberg chain the derivation does not apply in this
form at all: the ``sites'' entering $C_{\rm diag}$ are product-basis (computational) configurations,
not single-particle spatial positions, exchangeability under the physical permutation group is broken
by the open-boundary chain geometry and by the fact that $S_i^z$ couples to a specific, not
interchangeable, field $h_i$, and consecutive many-body eigenstates are correlated by the extensive,
not intensive, structure of the interacting Hamiltonian in a way the single-particle random-matrix
argument does not capture. We therefore present Eq.~(\ref{eq:selfavg}) as a controlled prediction, now tested for the
random-matrix models (up to the eigenstate-correlation correction $\alpha<1$ identified above), and only
as qualitative motivation -- not a derivation -- for the direction (suppression, not necessarily its
functional form) of the Heisenberg thermal washout.

\end{document}